\documentclass{article}
\usepackage{arxiv}
\usepackage{graphicx}
\pdfoutput=1
\usepackage[singlelinecheck=on, justification=raggedright]{caption}
\usepackage[labelfont=bf,textfont=normalfont,singlelinecheck=off,
justification=raggedright]{subcaption}
\usepackage{longtable}
\usepackage{algorithm}
\usepackage{algorithmic}
\usepackage{multirow}
\usepackage{makecell}
\usepackage{booktabs}
\usepackage[utf8]{inputenc} 
\usepackage[T1]{fontenc}    
\usepackage{hyperref}       
\usepackage{url}            
\usepackage{amsfonts}       
\usepackage{nicefrac}       
\usepackage{microtype}      
\usepackage{lipsum}
\usepackage{enumitem}
\usepackage{pifont}

\title{DNNVM : End-to-End Compiler Leveraging Heterogeneous Optimizations on FPGA-based CNN Accelerators\thanks{This paper appears in: IEEE Transactions on Computer-Aided Design of Integrated Circuits and Systems (TCAD). Manuscript received by February 19th, 2019; revised by April 13rd, 2019; accepted by June 30th, 2019; date of current version is July 16th, 2019. This work was supported by Xilinx and Beijing Innovation Center for Future Chips. This work was supported by Tsinghua Xilinx AI Research Fund, Beijing National Research Center for Information Science and Technology (BNRist), National Key R\&D Program of China 2018YFB0105005, National Natural Science Foundation of China(No. 61622403, 61621091), the project of Tsinghua University and Toyota Joint Research Center for AI Technology of Automated Vehicle(TT2018-01).}~~\thanks{Copyright (c) 2015 IEEE. Personal use of this material is permitted. However, permission to use this material for any other purposes must be obtained from the IEEE by sending an email to pubs-permissions@ieee.org.}}

\author{Yu Xing\thanks{Y. Xing, L. Sui, X. Jia, X. Liu, Y. S. Wang, Y. Shan are with Xilinx, Beijing 100083, China.}~~\thanks{Y. Xing, S. Liang, J. Qiu, Y. Wang are with the Department of Electronic Engineering, Tsinghua National Laboratory for Information Science and Technology, Tsinghua University, Beijing 100084, China.}~~\thanks{Y. Xing, S. Liang, J. Qiu, Y. Wang are with Beijing National Research Center for Information Science and Technology (BNRist).}, ~Shuang~Liang\footnotemark[4]~~\footnotemark[5], ~Lingzhi~Sui\footnotemark[3]~, ~Xijie~Jia\footnotemark[3], ~Jiantao~Qiu\footnotemark[4]~~\footnotemark[5], ~Xin~Liu\footnotemark[3], ~Yushun~Wang\footnotemark[3], \\
\textbf{Yi~Shan\footnotemark[3], ~and ~Yu~Wang\footnotemark[4]~~\footnotemark[5]} \\
\texttt{yuxing@xilinx.com, yu-wang@tsinghua.edu.cn} \\
}

\begin{document}

\maketitle

\begin{abstract}
The convolutional neural network (CNN) has become a state-of-the-art method for several artificial intelligence domains in recent years. The increasingly complex CNN models are both computation-bound and I/O-bound. FPGA-based accelerators driven by custom instruction set architecture (ISA) achieve a balance between generality and efficiency, but there is much on them left to be optimized. We propose the full-stack compiler DNNVM, which is an integration of optimizers for graphs, loops and data layouts, and an assembler, a runtime supporter and a validation environment. The DNNVM works in the context of deep learning frameworks and transforms CNN models into the directed acyclic graph: XGraph. Based on XGraph, we transform the optimization challenges for both the data layout and pipeline into graph-level problems. DNNVM enumerates all potentially profitable fusion opportunities by a heuristic subgraph isomorphism algorithm to leverage pipeline and data layout optimizations, and searches for the best choice of execution strategies of the whole computing graph. On the Xilinx ZU2 @330 MHz and ZU9 @330 MHz, we achieve equivalently state-of-the-art performance on our benchmarks by na\"ive implementations without optimizations, and the throughput is further improved up to 1.26x by leveraging heterogeneous optimizations in DNNVM. Finally, with ZU9 @330 MHz, we achieve state-of-the-art performance for VGG and ResNet50. We achieve a throughput of 2.82 TOPs/s and an energy efficiency of 123.7 GOPs/s/W for VGG. Additionally, we achieve 1.38 TOPs/s for ResNet50 and 1.41 TOPs/s for GoogleNet.
\end{abstract}


\section{Introduction}
Deep convolutional neural networks (CNNs) \cite{vgg, resnet, googlenet, mobilenet} are extensively employed in various artificial intelligence tasks, such as object detection, classification, natural language processing and semantic segmentation. The extensive variety in application complexity produces substantial challenges for hardware platforms \cite{Chen2017Platform}, such as computation ability and power consumption.

Multi-core CPUs and GPUs have been the dominant hardware platforms for CNN training and inference. Following the single-instruction, multiple-data (SIMD) or single-instruction, multiple-thread (SIMT) parallel-processing methods, CNN algorithms can be efficiently processed. While the potentialities of a CPU have been fully exploited\cite{optimizecpu}, unfortunately, a CPU cannot provide acceptable computation ability for CNN models. The high utilization of a GPU relies on a large batch size, which indicates that input feature maps are processed in parallel; thus, GPUs are very suitable for training. However, the applications of CNN in practice, especially vision tasks and video processing, require input feature maps to be separately executed. The low utilization of GPU resources in inference, high cost and relatively low energy efficiency limit the applications of GPUs for CNNs.

There is a significant trend of selecting custom hardware platforms, such as field-programmable gate arrays (FPGAs) and application-specific integrated circuits (ASICs), as the next-generation of CNN accelerators for inference. ASICs such as TPU\cite{tpu}, Nervana\cite{nervana} can achieve state-of-the-art performance; however, the design of ASICs is time-consuming and expensive to compete with the rapid evolution of CNN algorithms. FPGA-based accelerators that target a specific CNN model achieve appealing performance while sacrificing the flexibility to different networks and platforms. To avoid this problem, some authors have proposed the alternative of hardware templates with many configurable parameters \cite{towards}. However, hardware designers are concerned with the low-level hardware behaviours in a cycle-accurate manner or optimize the frequency and throughput by a long process of parameter value selection. Although the use of high-level implementation tools such as Vivado High-level Synthesis (HLS) can simplify this design process, it is still a tedious task to describe an efficient parallel architecture through long compilation to improve throughput by HLS.

An alternative solution is to construct a flexible hardware structure with a compiler to map different CNN models onto it by generating different instructions. In this way, we can take advantage of both the software programmability and the efficiency of custom hardware. Additionally, several advanced compiler technologies, graph-level algorithms, optimization methods can be applied in a compiler to improve the throughput of accelerators and the productivity. Building on these considerations and challenges, we briefly introduce our hardware architecture extended from Angel-Eye\cite{angeleye} and propose a full-stack compiler infrastructure, our main contributions are described as follows:

\begin{itemize}[leftmargin=*]
\item An end-to-end compilation infrastructure named the Deep Neural Network Virtual Machine (DNNVM) is proposed for a custom hardware design that is extended from angel-eye. The DNNVM is an integration of optimizers for graphs, loops and data layouts, an assembler, a runtime supporter and a validation environment.

\item In the DNNVM, we employ a domain-specific computing graph named XGraph to decouple the DNNVM with various deep learning frameworks and hardware platforms. Based on XGraph, we propose an efficient subgraph isomorphism algorithm to enumerate all valid potentially profitable fusion opportunities and optimize the pipeline.

\item We adopt a heuristic shortest-path algorithm to extend the optimization scope to the whole computing graph and obtain the best choice of fusion strategies, which is primarily disregarded by other designs.

\item Compared with na\"ive implementations without fusion, the experimental results demonstrate up to 1.26x improvement on our benchmarks VGG \cite{vgg}, ResNet \cite{resnet}, and GoogLeNet \cite{googlenet}. We achieve the state-of-the-art performance on ZU2 for embedded applications, and on ZU9 for VGG and ResNet.

\end{itemize}

The remainder of this paper is organized as follows: Section 2 introduces the background and motivations of this research. An overview of DNNVM and our custom hardware is provided in Section 3. We leverage a heuristics algorithm to explore pipeline optimizations in Section 4 and apply a heuristics shortest-path algorithm to obtain the optimal execution strategy in Section 5. The performance that we achieve is presented in Section 6. Section 7 introduces related studies about optimizations in hardware designs and implementations of compiler designs. We conclude this paper in Section 8.

\section{Background and Motivations}

\subsection{CNN Accelerators}

Accelerators for CNNs can be classified into two categories: (1) general-purpose processors (GPP), such as CPUs \cite{intel} and GPUs \cite{nvidia}, (2) specialized domain accelerators (SDAs) that are predominantly implemented on FPGAs \cite{angeleye, dnnweaver, xfdnn, fpgaconvnet17, vta} or ASICs \cite{cambricon}. GPPs are concerned with flexibility, while SDAs can offer 10-100$\times$ more energy efficiency than GPPs in dedicated application domains. Due to the advantages of reconfigurability, flexibility and energy-efficiency, FPGA-based SDAs have attracted a considerable amount of attention. Some FPGA-based hardware designs like fpgaConvNet\cite{fpgaconvnet17} and DeepBurning\cite{deepburning} favour customization over generality, by using high-level synthesis (HLS)\cite{opencl}, they can directly map each layer or subgraph of a target CNN model into one computation block in an elegant way. However, each block needs to be designed and parametrised carefully and FPGA needs to be reconfigured for each input CNN model. In this study, we employ a design trade-off between generality and customization by introducing a custom instruction-set architecture (ISA)\cite{angeleye, snowflakecompiler, dla}, our hardware architectures are easily scaled and configured based on the availability of hardware resources. The corresponding compiler are designed for better throughput. Our design not only guarantees compatibility with neural networks with different scales and topologies but also achieves appealing performance. Without reconfiguring the FPGA, multiple CNNs can be consecutively implemented by running different instructions generated by our compiler.

\subsection{Compiler Toolchains}

A compiler is generally an architecture-aware tool that efficiently maps algorithms to hardware implementations and control signals. The generation of CPU instructions from CNN algorithms is relatively simple due to advanced compiler technologies and a mature compiler ecosystem, such as LLVM \cite{llvm}. Additionally, explorations in data scheduling optimizations \cite{fusetile}, such as fusion, tiling, vectorizing, and parallelism \cite{halide}, and many other affine transformation methods \cite{polymage}, are utilized to produce instructions for efficient CNN processing. Algorithms that are capable of auto-tuning and optimizing scheduling attach a substantial amount of attention as compiler optimization methods \cite{autoschedule}.

Compilers for FPGA-based accelerators or ASICs are responsible for tuning the configurable parameters of the hardware architectures \cite{dnnweaver,dla}, and map an NN to custom instructions \cite{snowflakecompiler,dla,xfdnn}. These domain-specific compilers maximize the hardware efficiency according to different topologies of CNN and hardware platforms on both the hardware side and the software side. Due to the immature ecosystem and originality of custom hardware designs, explicit scheduling and memory management are required for each compiler that targets custom hardware platforms. In the literature, few compilers for custom FPGA-based accelerators, such as TVM \cite{tvm}, DLA\cite{dla}, have been addressed as blueprints for standardized compiler tool chains. Inspired by these studies and traditional compilers \cite{dlvm, tensorcomprehension, halide, tvm} for GPPs, we present a complete end-to-end compiler toolchain for our custom hardware design.

\subsection{Optimization Methods}

\begin{figure*}
\centering
\includegraphics[width=0.98\textwidth]{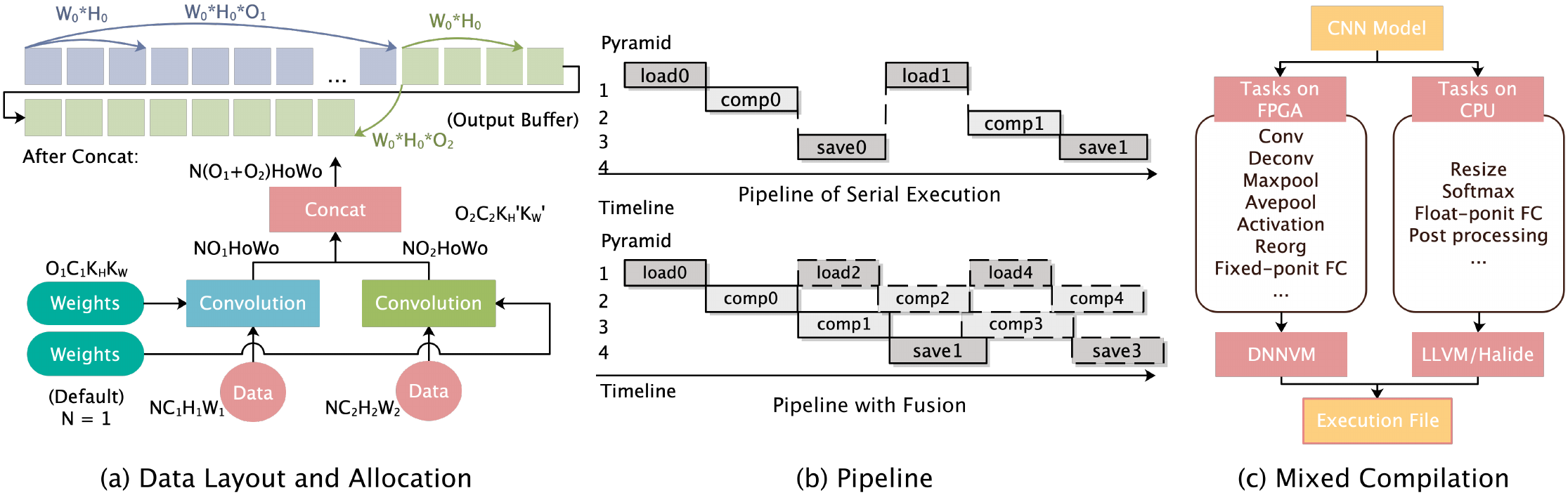}
\caption{Motivations of Compiler Optimization Methods, (a) data layout in Caffe, $N$ means batch, $NCHW$ is for feature maps, width of a feature map is first and height is the next, $OCK_{h}K_{w}$ is for weights, kernel is first and channel is next, (b) operation fusion allows concurrent implementations of computation and optimizes pipeline, (c) automatically distribute tasks to CPU and FPGA.}
\label{fig:motivation}
\end{figure*}

\subsubsection{Intermediate Representations(IR)}

The first step of a compiler is transforming a high-level framework-dependent CNN representation to an IR. For example, Caffe\cite{caffe}, Darknet\cite{darknet} adopt layer-based IRs, which define each typical CNN operation as a layer. A specific layer can be optimized in a straightforward manner to produce an efficient implementation. Unfortunately, difficulties arise when implementing cross-layer optimizations by layer-based IR. TensorFlow \cite{tensorflow} employs a graph-level IR and constructs computing graphs. Each node in a computing graph represents a course-grained operation, such as convolution, or a fine-grained operation, such as add and pad. Various graph algorithms\cite{tvm} can be leveraged to optimize the implementations. Due to the difference from different frameworks such as the granularity of operations and the boundary conditions. To optimize operator implementations from different deep learning frameworks, we need to make $O(N_f \cdot N_o \cdot N_p)$ efforts, where $N_f$ is the number of deep learning frameworks, $N_o$ is the number of operations, and $N_p$ is the number of platforms. To reduce the complexity of optimizations, an appropriate IR to decouple the compiler with deep learning frameworks and hardware platforms is necessary.

\begin{table}[htbp]
\setlength{\abovecaptionskip}{0em}
  \tiny
  \centering
  \caption{The Symbol List}
  \setlength{\tabcolsep}{0.6mm}{
  \renewcommand{\arraystretch}{0.8}
  \begin{tabular}{c|c|c}
    \toprule \\[-8pt]
    \thead{Inputs} & \thead{Section} & \thead{Comment}\\[-3pt]
    	\hline \\[-7pt]
    \thead{W, H, C}  & \thead{2, 3, 4} & \thead{Width, height and channel of feature maps}\\[-3pt]
        \hline \\[-7pt]
    \thead{N}  & \thead{2, 3} & \thead{Batch of feature maps, default = 1}\\[-3pt]
        \hline \\[-7pt]
    \thead{$K_{w}$, $K_{h}$, O}  & \thead{2, 3, 4} & \thead{Width, height and number of kernels}\\[-3pt]
        \hline \\[-7pt]
    \thead{IC, OC}  & \thead{3, 4} & \thead{Input / Output channel of an operation}\\[-3pt]
        \hline \\[-7pt]
    \thead{h\_p,inc\_p,oc\_p} & \thead{2, 4} & \thead{Parallelism upon dimension of H, IC or OC.}\\[-3pt]
        \hline \\[-7pt]
    \thead{$A_{comp}$, $A_{ac}$}  & \thead{4} & \thead{Amount of computations and data exchange}\\[-3pt]
        \hline \\[-7pt]
    \thead{$F^{-1}(W)$}  & \thead{4} & \thead{Corresponding width of an input feature map}\\[-3pt]
        \hline \\[-7pt]
    \thead{$G^{-1}(H)$}  & \thead{4} & \thead{Corresponding height of an input feature map}\\[-3pt]
        \hline \\[-7pt]
    \thead{$T_{\_(i)}$}  & \thead{4} & \thead{Tile size upon one dimension of the $i^{th}$ operation}\\[-3pt]
    \bottomrule
    \end{tabular}}
  \label{tab:symbol}%
\end{table}%

\subsubsection{Data Layout and Allocation}

Mapping data onto on-chip buffers helps enhance data locality and improve the total throughput. An implementation of operations usually requires a specific data layout of input data. In Figure \ref{fig:motivation}(a), the data layout of feature maps in Caffe\cite{caffe} is NCHW. Additionally, it is a common practice to partition an operation into pieces to parallelize among multiple cores in a CPU and blocks in a GPU. For example, instead of the default (NCHW or NHWC) from Caffe or Tensorflow, the data layout of NCHW[x]c is adopted for CONVs in an optimized implementation\cite{optimizecpu} on a CPU , in which c is a sub-dimension of C, and the number x indicates the size of the sub-dimension($channels = sizeof(C)\times sizeof(c), x = sizeof(c)$). Besides, as the resolution of an image increases and the neural networks become deeper, the size of on-chip buffers is not sufficient to store all required data. The feature maps also need to be tiled and re-organized. But transformations of data layout and dimension-related operations such as flatten, concat and reorganization, introduce significant overheads. As a result, a specific data layout mode for our specialized accelerator and an appropriate data slicing rule to allocate on-chip memory for each tiled data should be considered.

\begin{figure*}
\includegraphics[width=0.95\textwidth]{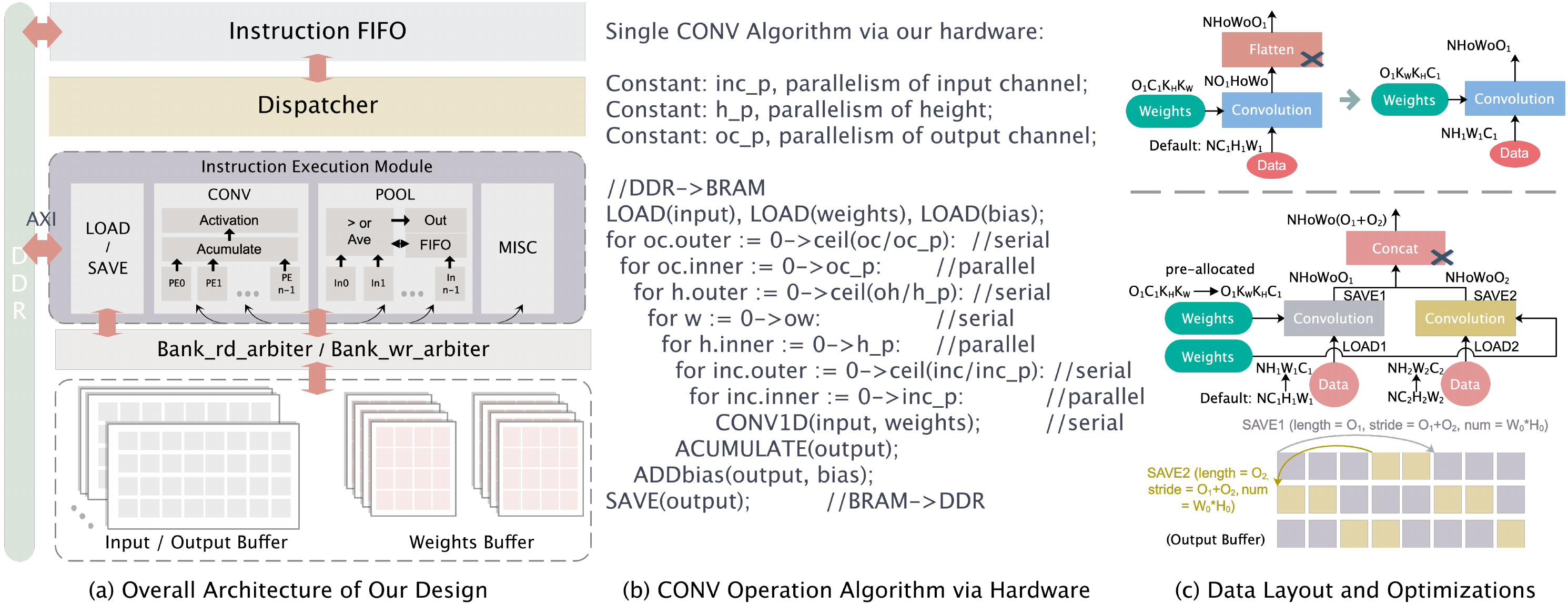}
\caption{(a) Overall architecture of our Design, (b) Implementation algorithms of CONV via our hardware, parallelisms are set upon input channel, input height and output channel, (c) the date layout of NHWC is set for feature maps, NWHC is set for weights, dimension transformation such as flatten and concat can be pruned and fused to SAVE of the previous operation.}
\label{fig:hardware}
\end{figure*}

\subsubsection{Pipeline}

As shown in Figure \ref{fig:motivation}(b), pipeline parallelism enables simultaneous implementations of computation and data communication in a single operation. Additionally, operation Fusion, which fuses adjacent operations, has been demonstrated as an effective pipelining technology in various hardware platforms \cite{tvm, fusetile, halide, tensorcomprehension, fuse1, fuse2}. In this way, on-chip input feature maps will perform fused operations to produce output feature maps without intermediate results for communication with the off-chip memory. Operations can be concurrently executed in parallel computation engines for latency hiding. Fusion technology has been employed in multiple specialized acceleration platforms, such as fpgaConvnet \cite{fpgaconvnet17} and VTA \cite{vta}. However, architectures such as fpgaConvnet leverage fusion on the hardware side. On the software side, compilers such as VTA can only fuse limited operations. We extend the optimization scope to the entire computing graph and propose heuristic algorithms instead of greedy algorithms to ensure the inclusion of several potentially profitable fusion strategies.

\subsubsection{Verification}

A CNN is trained with 32-bit floating point data on a GPU or CPU, and data are usually simplified to a fixed-point format with the same bit-width such as 8-bit\cite{angeleye} or 2-bit\cite{2bit}, with negligible accuracy loss to reduce on-chip memory consumption and computation complexity. The hardware design should be adapted to the fixed-point computation. On the software side, we need to guarantee that the calculated results achieved through various optimization methods in a compiler are consistent with the results achieved by hardware. However, the verification method is primarily disregarded in previous research.

\subsubsection{Mixed Compilation}

State-of-the-art CNNs use a variety of algorithms and technologies to solve complex artificial intelligence tasks. Implementing all operations on FPGAs may be inefficient and infeasible, which increases the design complexity of accelerators and is time-consuming. Computationally intensive operations are accelerated on FPGA while other operations can be implemented by a CPU. To simplify the usability of the tool chain, a compiler should automatically distribute different operations to a CPU and FPGA, generate instructions for specialized domain accelerators and produce CPU code by general compilers.

Based on these considerations, we leverage several optimization methods in DNNVM for a well-designed hardware architecture, and greatly improve the productivity and the overall throughput of the system.

\section{Framework Overview}
This section provides an overview of our hardware design and the DNNVM stack.

\subsection{Hardware Design}

DNNVM's instruction set architecture (ISA) is composed of four kinds of instructions: LOAD/SAVE, CONV, POOL, and MISC. Execution modules are designed to correspond to our customized ISA. The implementations of computation modules are inspired by Angel-Eye \cite{angeleye}. \textbf{LOAD/SAVE} modules move data between on-chip buffers and off-chip DDR. The \textbf{CONV} module operates convolution over its input data that are fetched from the input buffers. The \textbf{POOL} module operates pooling over its input data, during which one bit is used to specify the pooling type. \textbf{MISC} modules execute other operations, such as element-wise add, reorganization, start, and end. These instructions are variable in size, and a few bits are set for the dependency relation among instructions to guarantee that they are executed in a certain way. The coarse-grained nature of the ISA enables the accelerators to incorporate graph-level optimizations and instruction pipelining, regardless of the overhead from decoding and fetching for a large number of low-level instructions.

\begin{figure*}
\includegraphics[width=0.98\textwidth]{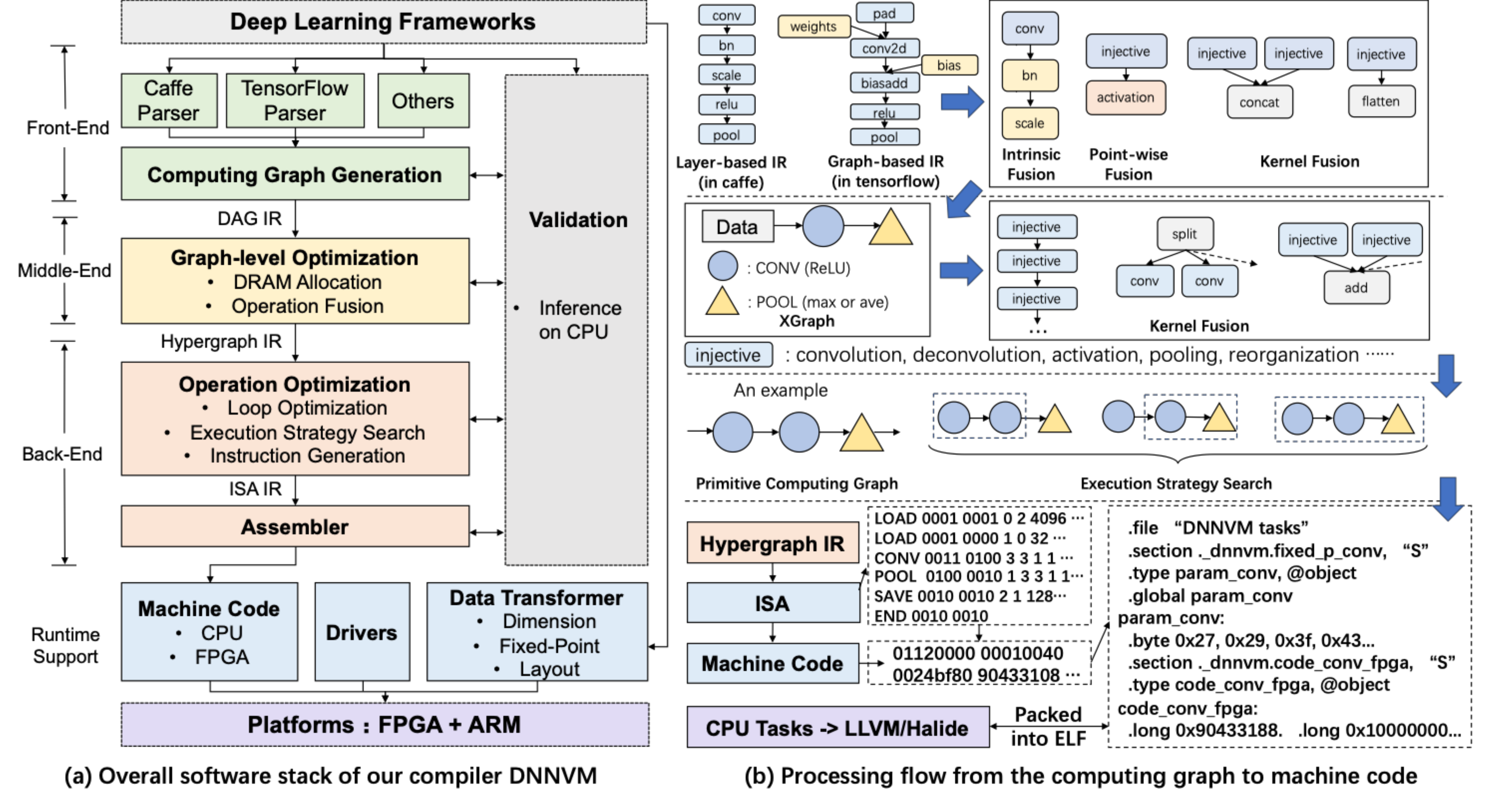}
\caption{(a) Overall software stack of our compiler DNNVM, (b) DNNVM transforms the computing graph into course-grained XGraph through intrinsic fusion, ponit-wise fusion and prunes all data transformation operations; then leverages various kernel fusion and searches for the best choice of fusion strategies, maps the algorithms into custom ISA and machine code, and finally packs the instructions for FPGA with CPU code.}
\label{fig:compiler}
\end{figure*}

Another important advantage of our ISA is that each computation module can be reused for various operations. For example, the convolution module can be reused for deconvolution, depth-wise convolution, and dilated convolution by reconfiguring fields such as stride, length, and mode in conv-related instructions.

As shown in Figure \ref{fig:hardware} (a), instructions are fetched from off-chip \textbf{DDR} to \textbf{Instruction FIFO}. The \textbf{Dispatcher} is responsible for decoding instructions into controlling signals and addresses of registers for the execution modules. The \textbf{Bank arbiter} is used to determine the order of access to buffers among requests. We use BRAMs and DSP slices to construct \textbf{Buffers} and ALUs and pack two or four INT8 Mults \cite{706, 490} into one DSP48E1 slice (25$\times$18 DSP configurations) to achieve the peak performance of 380 GOPs/s and 4.05 TOPs/s on ZU2 devices and ZU9 devices, respectively at 330 MHz. A fixed number of BRAMs are pre-allocated for input feature maps, output feature maps and parameters. Each buffer can be reused by instructions with different access addresses. In Figure \ref{fig:hardware} (b), we partition CONV into several pieces and parallelize upon input channel, height and output channel. In this way, we do not need to leverage optimizations for specific kernel sizes and enhance the flexibility. The algorithm in Figure \ref{fig:hardware} (b) decides the data layout of feature maps to be NHWC, which means input channel first, and batch is the last. The data layout of OWHC is for weights, and they can be pre-allocated on DDR before computation so that no extra overhead would be introduced. In Figure \ref{fig:hardware} (c), different data transformation operations such as flatten and concat may introduce significant overheads. Based on our understanding of the hardware design and computation implementations, our custom data layout method prunes the flatten operation naturally due to the completely same layout of the results. Other dimension transformation operations can be fused to the SAVE instruction of the previous operation. For example in Figure \ref{fig:hardware} (c), a channel-first data layout makes SAVE1 stores the first $O_{1}$ pixels at the address 0, and then strides $O_{1}+O_{2}$ pixels to store the second $O_{1}$ feature maps in the output buffer, which is equivalent to a single concat after convolution.

\subsection{Compiler Infrastructure}

The DNNVM is a full-stack compiler that takes advantage of many advanced compiler technologies \cite{tvm,tensorcomprehension,dla}. As presented in Figure \ref{fig:compiler}, the DNNVM transforms CNN models into a framework-independent computing graph, searches for pipeline optimization opportunities, allocates buffers for fused-operations, and importantly, selects the best choice of execution strategies. In general, DNNVM integrates various optimizers, runtime support, a data transformer and a validation bench to improve the productivity.

\begin{figure*}
\setlength{\abovecaptionskip}{0em}
\includegraphics[width=0.96\textwidth]{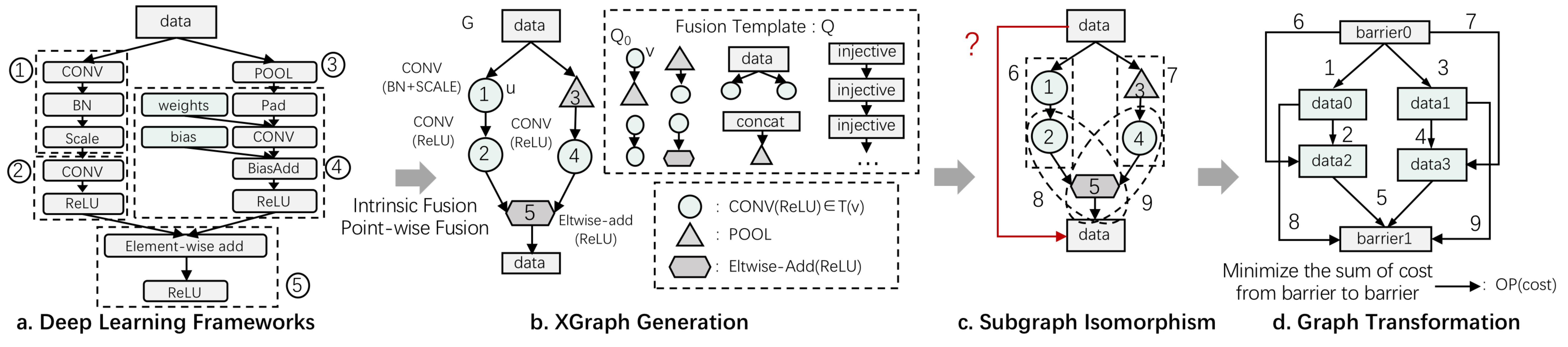}
\caption{A case study of the operation fusion. DNNVM works in the context of deep learning frameworks and transform CNN models into XGraph by leveraging intrinsic fusion and point-wise fusion, then searches for fusion opportunities according to pre-defined fusion templates by a subgraph isomorphism algorithm.}
\label{fig:processflow}
\end{figure*}

\begin{itemize}[leftmargin=*]

\item \textbf{Front-End and XGraph:}
DNNVM presents the coarse-grained computing graph format XGraph to decouple the compiler with deep learning frameworks and hardware platforms to reduce the complexity to $O(N_o)$. As shown in Figure \ref{fig:processflow} and Figure \ref{fig:compiler} (b), different deep learning frameworks have a great diversity of operations with different granularities, such as \textcircled{1} from caffe and \textcircled{4} from TensorFlow in Figure \ref{fig:processflow}. As shown in the second row in Figure \ref{fig:compiler} (b), XGraph has a similar format with Caffe and Pytorch but is a graph-level IR. DNNVM transforms the computing graphs from different deep learning frameworks into the course-grained XGraph by leveraging intrinsic fusion such as the convolution + BN + Scale, which can be pre-calculated, and point-wise fusion, such as convolution + ReLU. Additionally, dimension transformation operations are pruned or fused into the previous computation operation as well.

\item \textbf{Middle-End:}
We recognize three types of operation fusion templates to maximize the utilization of locality and parallelism and optimize the pipeline. In the middle-end of DNNVM, we adopt graph-level optimizations. A heuristics subgraph isomorphism algorithm is designed to efficiently enumerate all fusion opportunities. In addition, using the directed acyclic graphs (DAG) of computational operations, we design an efficient configuration parser based on a programming language named scheme \cite{scheme} for these hypergraph IRs and automate the DRAM allocation, synchronization, and distribution according to these IRs. We primarily allocate DRAM for feature maps and parameters. The space allocated for the feature maps which are not depended by the following operations can be freed, while the space for parameters such as weights and bias are protected.

\item \textbf{Back-End:}
In the back-end, we leverage a combination of affine transformations, such as hierarchical tiling using nested loops and pipelines. We search for the best choice of execution strategies by a heuristics shortest-path algorithm to maximize the utilization of locality and parallelism. Then we map this execution strategy into our ISA IRs. Assembler transforms the instructions into the machine code.

\item \textbf{Runtime Support:}
This module implements the execution instructions, built-in functions, and other fundamental behaviours of applications. Computations which are not supported by our hardware design are compiled by LLVM \cite{llvm}, or Halide \cite{halide} for acceleration. Machine code, the reshaped fixed-point parameters, CPU code and float parameters are integrated in an Executable and Linking Format (ELF) file. The executable file using hardware drivers achieve total control over the hardware platform for various applications.

\item \textbf{Validation Module:} First, at the beginning of the hardware design, we need to make a trade-off between the accuracy loss and the hardware resources consumption to determine the shifting, truncation, and rounding methods. So we add specific functions to describe these modifications according to our hardware design. Second, for float point data, underlying libraries and the order of execution adopted by different deep learning frameworks influence the final results\cite{paddlepaddle}. But in our design, all data in neural network is set to the fixed-point format with the same bit-width, and the execution order will not change the final results. We re-implement operations from different frameworks in fixed-point format and summarize the data layout and boundary condition of these operations. As a result, our validation tools can prevent even a one-bit difference between the results by the CPU and the results by the FPGA.

\end{itemize}

These modules are indispensable parts in a full-stack compiler design. We focus on the optimization methods in DNNVM in the following sections, especially the pipeline optimization and operation fusion of the entire computing graph.

\section{Optimizations}

As mentioned above, DNNVM works in the context of deep learning frameworks and transforms the computing graph of CNN model into XGraph. Prior optimization approaches \cite{optimize} usually consider each individual computation operation separately and start with the assumption that each operation loads the feature maps and parameters from the off-chip memory and then writes the intermediate results back to the off-chip memory. We make $A_{comp}, A_{ac}$ denote the amount of computations and data exchange for each operation. We extend Computation to Communication (CTC) \cite{optimize}, which describes the computation operations per memory access to describe the influence of fusion. The total amount of computation is fixed when given a CNN, and the performance improvement is in proportion to the reduction in communication.

When given a CNN model with $l$ operations. In previous studies, without operation fusion techniques, CTC would be:
\begin{equation}
  CTC = \frac{\sum_{i=0}^{l} A_{comp}}{\sum_{i=0}^{l} A_{ac}}
  \label{eq:ctcnof}
\end{equation}

Ideally, the buffers are sized to fit the entire feature maps and parameters. With the operation fusion:
\begin{equation}
  CTC = \frac{\sum_{i=0}^{l} A_{comp}}{\sum_{i=0}^{l} A_{ac} - \sum_{i=1}^{l-1} A_{ac}}
  \label{eq:ctcf}
\end{equation}

Unfortunately, on-chip memory is often too small to store the required data of fused-operations, even is too small for a single operation. For example, VGG has 16 convolution layers with more than 100 MB parameters; however, Xilinx ZU2 has only 0.66 MB of on-chip BRAMs. To overcome this constraint, segmenting the CNN models into subgraphs is a potential solution. As a result, \textbf{the first condition to be satisfied for fusion is that all feature maps and parameters required for a subgraph without tiling can be stored on-chip}. In this case, we determine that more than two operations can be fused; the number of operations to be fused is not the limitation. Our heuristics start from each operation and then iterate over the computing graph to check for fusing opportunities.

Besides, we find that longitudinally adjacent operations such as Conv + Conv, can avoid frequent data exchange with the off-chip memory and reduce the bandwidth requirement. The time consumed for fused-operations such as Conv + Pool can be greatly reduced by the concurrent execution of communication and different computation blocks. Horizontally adjacent operations share the same input feature maps, which can be reused by different operations without reloading. Additionally, due to the course-grained attributes of XGraph, the generation process of XGraph can be seen as operation fusion. Fined-grained operations are fused into a complex operation. Dimension transformation operations such as Flatten and Concat can be pruned or fused into the previous operation. 

Building on these considerations, we transform the XGraph generation and pipeline optimization into an operation fusion problem. We build a set of fusion templates, \textbf{the second condition for fusion is that dependency among operations is pre-defined in our fusion templates} shown in Figure \ref{fig:compiler}. Operations are fused when either of these two conditions is satisfied. We introduce the fusion templates in subsection A and present the our operation fusion opportunities traversal algorithm in subsection B.

\begin{figure*}
\setlength{\abovecaptionskip}{0em}
\includegraphics[width=0.98\textwidth]{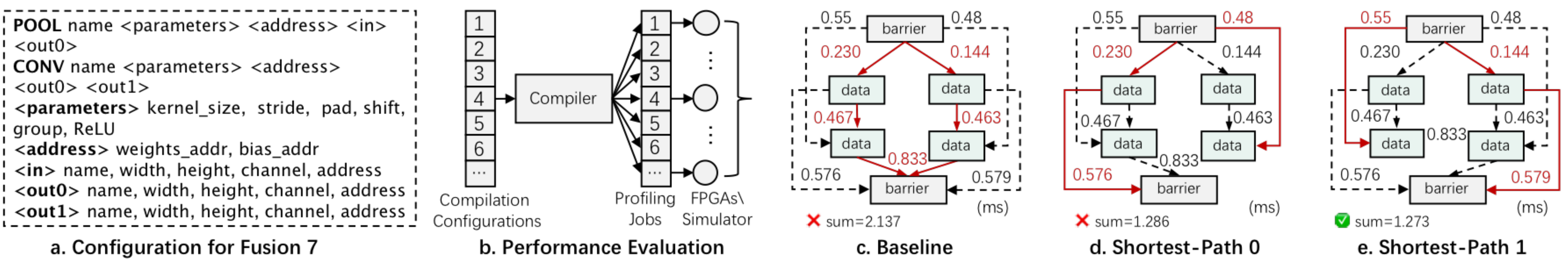}
\caption{Evaluation and path searching. By resolving configurations generated by DNNVM in Scheme, DNNVM automatically evaluates implementation costs on-board or by simulator, and searches for the best choice of execution strategies by a heuristic algorithm.}
\label{fig:path-search}
\end{figure*}

\subsection{Fusion Template}

We recognize three categories of the operation fusion, as shown in right column of Figure \ref{fig:compiler}: intrinsic fusion, point-wise fusion and kernel fusion. The $injective$ in templates can be convolution, pooling, nonlinear, deconvolution, depth-wise convolution, upsample, and reorganization in DNNVM.

\subsubsection{Intrinsic Fusion}

We fuse the parts of the graph that can be pre-computed or statically determined. For example, the parameters of Batch Normalization and Scale can be pre-computed, and these operations can be fused to the adjacent convolution. In addition, we fuse operations from different deep learning frameworks to a uniform grain. As shown in Figure \ref{fig:processflow}, we fuse pad, weights, bias, conv2d, biasadd and relu from TensorFlow to a single convolution operation in XGraph.

\subsubsection{Point-wise Fusion}

We fully integrate the point-wise operations. A typical example is the relu after convolution. We set one-bit which denotes the nonlinear type in the instruction CONV. Each intermediate result achieved by convolution will be sent to the nonlinear module directly to avoid the kernel boot time and to reduce the time required for off-chip memory access.

\subsubsection{Kernel Fusion}

We leverage kernel fusion for data reuse. First, horizontally adjacent layers share the same input feature maps; as a result, input data does not need to be reloaded. Second, vertically fused-operations can avoid the exchange of intermediate results between on-chip buffers and off-chip memory. Third, different operations can be concurrently executed by different computation modules. In addition, dimension transformation and the reorganization of feature maps can be skipped by fusing operations of reshaping, such as concat and flatten, into the SAVE process of the previous operator or the LOAD process of the following operator.

Once the DNNVM obtained a computation graph from the deep learning frameworks, it needs to identify all possibilities of operation fusion. We leverage a subgraph isomorphism algorithm for fusion opportunities traversal.

\subsection{Subgraph Isomorphism Algorithm}

Use of the fusion template for traversal matching on the computation graph represented by XGraph is a subgraph isomorphism problem, which is a well-known NP-complete problem. Currently, operation fusion techniques on GPPs employ a greedy method for fusion template matching. However, if we greedily search for fusion opportunities, several valid and potentially profitable opportunities may be missed. The sequence of template searching influences the final performance. For example, as shown in Figure \ref{fig:processflow} b, if we fuse \textcircled{4} with \textcircled{3}, the combination of \textcircled{4} and \textcircled{5} will be missed.

Fortunately, due to the coarse-grained definition of XGraph, the searching space is not too large and there are only hundreds of nodes in a computing graph. So we adopt a subgraph isomorphism algorithm to enumerate all fusion opportunities instead of greedy algorithms. Ullmann \cite{Ullmann}, vf2 \cite{vf2}, and boostIso \cite{boostiso} present several methods for subgraph isomorphism, and Jinsoo et al. \cite{depthcomparison} compare their performances. We learn from these methods and apply a heuristic algorithm for our design:

\begin{algorithm}[htb]
\caption{Subgraph Isomorphism Algorithm}
\begin{algorithmic}[1]
\renewcommand{\algorithmicrequire}{\textbf{Input:}}
\renewcommand{\algorithmicensure}{\textbf{Output:}}
\REQUIRE ~~Subgraph pattern template $Q, Q_i \in Q$;\\
\REQUIRE ~~computing graph $G$, start point $S_{i}$;\\
\ENSURE ~~all Subgraph isomorphisms of $Q_{i}$ in $G$;\\
\STATE $M := \emptyset$;
\FOR{each $v \in V(Q_{i}$)}
\STATE $C(v) := FilterCandiates(Q_{i}, G, v)$;
\IF{C(v) == $\emptyset$}
\RETURN;\\
\ENDIF;\\
\ENDFOR
\STATE $S_{i} := DefineStartPoint(Q_{i}, G)$\\
\STATE $SubgraphSearch(Q_{i}, G, M, S_{i}, e, v, \dots$);\\
\STATE \textbf{DEF} $SubgraphSearch()$:\\
\IF{$|M| == |V(Q_{i})|$}
\STATE     \textbf{Return} $M$;\\
\ELSE
\STATE $v := NextQueryVertex(S_{i}, Q_{i}$);\\
\STATE $C_{v} := RefineCandidates(v, C_{v}, G)$;\\
\FOR{each $u \in C_{v}$}
\IF{$Matching(u, v, \dots$)}
\STATE $UpdateState(M, v, u, \dots$)\\
\STATE $SubgraphSearch(Q_{i}, G, M, v, \dots$)\\
\ENDIF
\ENDFOR
\ENDIF \\
\end{algorithmic}
\label{alg:SIM}
\end{algorithm}

\begin{figure*}
\includegraphics[width=1\textwidth]{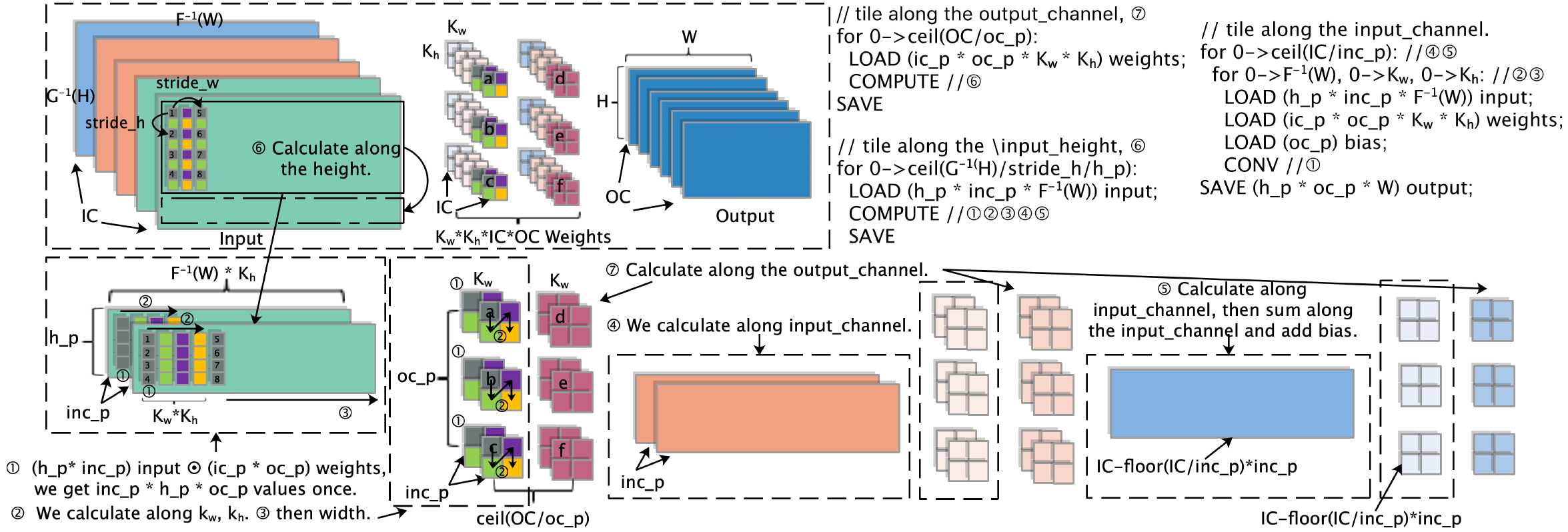}
\caption{Details of tiling for a single convolution. The dimension of tile is set along input channel (\ding{175} and \ding{176}), height(\ding{177}) and output channel (\ding{178}). Parallelism = inc\_p * h\_p * oc\_p (\ding{172}). Implementations along kernel (\ding{173}) and width (\ding{174}) are serial.}
\label{fig:tile details}
\end{figure*}

The computing graph of the CNN model $G$ is defined as a $<U, E, T>$, where $U$ is a set of vertexes, each vertex is a function that represents an operation in neural networks, $E$ represents the edges which represent the dataflow dependency and $T$ is a labelling function that maps the type and configurations of one operation. We use the symbol $Q$ to denote the set of fusion templates. $Q_i$ $\in$ Q, $Q_i$ represents one candidate fusion template. $v$ and $b$ are the symbols for a query vertex and a query edge in one fusion template. The problem is to identify all distinct embeddings of $Q_i$ in $G$.

In Algorithm \ref{alg:SIM}, line 2-7 $FilterCandiates$ searches for candidate vertexes $C(v)$ in G that have the same type of $v$ in $Q_{i}$ ($[T(v)==T(C(v)) | C(v) \in G]$). There are several vertexes in a fusion template $Q_{i}$, so we need to determine the start point $S_{i}$ in line 8. For example, the input computing graph $G$ has 120 convolutions and 15 poolings. The fusion template is Conv + Pool. If we set the Conv as the start point, we need to explore the adjacent vertexes of 120 convolutions, if we set the Pool as the start point, only 15 poolings' adjacent vertexes need to be explored. Here we set the start point $S_{i}$ to a type in $Q_{i}$ that has the least number of occurrences to reduce the size of the recursive call tree. In line 14-15, we select a vertex adjacent to the previously matched query vertex $M$ by $NextQueryVertex$, a breadth-first search is implemented here. By using $RefineCandidates$, we prune any vertex $u$ in $C_{v}$ such that $u$ is not adjacent to the previously matched vertex. $u$ would not be pushed back into $M$ (line 17-18) unless $u$ satisfies all query requirements, such as the type of the operation, the kernel size, and the stride. The edge that connects $u$ with a previously matched vertex should have the same type with a corresponding edge in the query. The recursion stops (line 11-12) when the algorithm obtains the complete solution (i.e., when $|M| = |V(Q_{i})|$). Otherwise, the algorithm calls $NextQueryVertex$ to select the query vertex $v \in V(Q_{i})$, which has not been matched.

\subsection{Tiling}

In the first condition for fusion as mentioned above, no intermediate results are communicated between DDR and on-chip buffer. However, the size of required input data and intermediate results among fused-operations achieved by fusion templates may exceed the capacity of on-chip memory. Even on-chip memory is not sufficient for a single operation. Slicing along the feature maps should be taken into consideration. As depicted in Figure \ref{fig:tile details}, we make $W$ and $H$ represent the width and height, respectively, of an output feature map, where $F^{-1}(W)$ and $G^{-1}(H)$ denote the corresponding width and height of an input feature map. We obtain the total amount of computations, which is a fixed number and can be calculated in Equation (\ref{eq:comp}) and the execution time can be depicted in Equation (\ref{eq:ac}):
\begin{equation}
  A_{comp} = 2 \times k_{w}\cdot k_{h}\cdot IC\cdot OC\cdot H\cdot W
  \label{eq:comp}
\end{equation}
\begin{equation}
  Time = \sum_{i=0}^{n}(LOAD_{i} + CONV_{i} + SAVE_{i})
  \label{eq:ac}
\end{equation}

$[LOAD_{i}, CONV_{i}, SAVE_{i}]$ represent the time used to load input and parameters, compute and save results for achieving each tiled output feature maps, respectively. $[T_{w}$, $T_{h}$, $T_{oc}]$ comprises a specific tile size combination for the width, height, channel of \textbf{output feature maps} respectively. $n$ denotes the number of tiles. We can find slicing causes reloading of feature maps and parameters between two neighbouring tiles. If n is large, the latency of reloading might be significantly unacceptable. As a result, we make the tiled size of the height and channel for output feature maps to be a fixed number and maximize the tiled size for output feature maps along the width according to Equation (\ref{eq:t}) and (\ref{eq:cons}). We can also extend this tiling rule for fused-operations with the constrait that the size of tiled input, output feature maps and all required parameters can not exceed the capacity of corresponding on-chip buffers $[B_{in}$, $B_{weights}$, $B_{out}]$, to determine the T$_{w}$.

\begin{equation}
  T_{h} = h\_p,~~T_{oc} = oc\_p,~~T_{ic} = inc\_p
  \label{eq:t}
\end{equation}
\begin{equation}
\left\{\begin{array}{l}
  T_{w} * T_{h} * T{oc} \leq B_{out}\\
  T_{ic} * K_{w} * K_{h} * T_{oc} \leq B_{weights}\\
  T_{ic} * F^{-1}(T_{w}) * T_{h} \leq B_{in}\\
\end{array}\right.
  \label{eq:cons}
\end{equation}

As shown in Figure \ref{fig:tile details}, we take a single convolution as an example. The input feature maps are tiled along the height and channel with $T_{h} = h\_p = 4$, $step_{h} = stride\_h = 2$ and $T_{ic} = inc\_p = 2$, the corresponding weights are tiled along input channel and output channel with $T_{ic} = ic\_p, T_{oc} = oc\_p$. In Figure \ref{fig:tile details}.\ding{172}, we need to load $h\_p * inc\_p * F^{-1}(W)$ input feature maps and $ic\_p * oc\_p$ weights. The point-wise multiplications between input feature maps and weights are implemented in parallel. In Figure \ref{fig:tile details}.\ding{173}, we load the same amount of input feature maps and weights by moving along the height then width with $length_{h} = kernel\_h, length_{w} = kernel\_w, step = 1$ and achieve the intermediate results. In \ding{174} and \ding{175}, we repeat this process along the width of input feature maps with $length = F^{-1}(W), step = stride\_w$ and the dimension of input\_channel with $length = IC, step = ic\_p$. The weights and feature maps which have been loaded on-chip can be reused and do not need to be reloaded. In \ding{176}, we sum the intermediate results along the input channel together and add the bias. We achieve $h\_p * oc\_p * W$ output feature maps at this moment. Similarly, we calculate along the dimension of height and output channel as shown in \ding{177} and \ding{178}. On-chip buffers storing feature maps and parameters which are not dependent by the following implementations can be freed and reused.

\section{Execution Path Searching}

As shown in Figure \ref{fig:processflow}.c, after we find all embeddings of $Q_{i}$ in $G$, one operation can be performed by different fusion templates. For example, \textcircled{2} can be fused with \textcircled{1} and \textcircled{5}, but \textcircled{2} needs to be implemented only once. Thus, we need to determine the which fusion strategies are better. Two challenges occur in this circumstance: 1) execution cost of each fused operation needs to be investigated; and 2) due to the complex topologies of neural networks, combinatorial explosion occurs in searching the best choice of execution strategies. To solve these challenges, we present the following methods.

\subsection{Fused-Operations Evaluation}

To determine the best choice of execution strategies with existing fusion templates, we should investigate the cost of each fused operation, which denotes the execution time. As shown in Figure \ref{fig:path-search}.a, after parsing configurations written in our DSL, we employ three methods to evaluate the performance of these fused-operations. First, we can directly execute each subgraph by our hardware platforms, which is the simplest evaluation method and costs less than 1s. Second, in the case of an offline evaluation, we leverage a model-driven method to estimate the cost. Instead of providing an empirical formula, we design a small neural network to fit the cost with a deviation between 5\% and 10\%. However, performance improvements of some optimizations like Figure \ref{fig:path-search}.d and e, are less than 5\% and we give up this option as a result. Third, we design a cycle-accurate simulator so that the compiler optimization and hardware design can be synchronously designed and optimized. We record the number of cycles consumption for each hardware block according to our hardware design, and we can calculate the number of cycles required for each instruction. Then we insert each instruction into a time wheel after analyzing the dependencies among them. Unfortunately, it takes lots of efforts to make a software totally consistent with the behaviours of the hardware, and it is hard to extract the debugging information in a time wheel.

\begin{table}[htbp]
  \tiny
  \centering
  \caption{Comparison among Evaluation Methods}
  \setlength{\tabcolsep}{2mm}{
  \renewcommand{\arraystretch}{0.8}
  \begin{tabular}{c|ccc}
    \toprule \\[-8pt]
    \thead{Method} & \thead{On-Board} & \thead{Model} & \thead{Simulator}\\[-2pt]
    	\hline \\[-7pt]
    \thead{Deviation} & \thead{0\%} & \thead{5-10\%} & \thead{0\%}\\[-2pt]
        \hline \\[-7pt]
    \thead{Time} & \thead{\textless 1s} & \thead{\textless 1min} & \thead{\textgreater 10min}\\[-2pt]
    \bottomrule
    \end{tabular}}
  \label{tab:compeva}%
\end{table}%

\subsection{Heuristic Shortest Execution Path}

Then we hope to obtain the execution strategies, with the smallest cost of performing all operations in the computing graph. First, we exchange the attributes of the node and the edge, as shown in Figure \ref{fig:processflow}.d. In this way, once operations can be fused, such as \textcircled{1} and \textcircled{2}, an additional edge \textcircled{6} is inserted into the computing graph. 

\begin{algorithm}[htb] 
\caption{ShortestPath Fuction}
\renewcommand{\algorithmicrequire}{\textbf{Input:}}
\renewcommand{\algorithmicensure}{\textbf{Output:}}
\label{alg:ShortestPath} 
\begin{algorithmic}[1]
\REQUIRE ~~computation graph $G$, $V \in G$, cost by evaluation
\ENSURE ~~the best choice of execution strategies with minimal cost
\FOR{each pair of adjacent barriers}
\STATE $s$ := start vertex
\STATE $d$ := destination vertex
\IF{have branch}
\FOR{each $branch_{i}$}
\STATE $cost[v_{j}][d] = \infty$ $\mid$ ($v_{j}$ points to $d$ in other branches)
\STATE $cost_{i}$ = ShortestPath($branch_{i}$)
\FOR{other $branch_{j}$}
\STATE $cost_{j}$ = ShortestPath(from s to $v_{j}$)
\ENDFOR
\ENDFOR
\STATE the best choice of strategy = path of min($cost_{i} + \sum{cost_{j}}$)
\ELSE
\STATE ShortestPath(between adjacent\ barriers)
\ENDIF
\ENDFOR
\STATE \textbf{DEF} ShortestPath():\\
\FOR{$(k\ = \ 0;\ k\ <\ |V|;\ k\ ++)$}
\FOR{$(i\ = \ 0;\ i\ <\ |V|;\ i\ ++)$}
\FOR{$(j\ = \ 0;\ j\ <\ |V|;\ j\ ++)$}
\STATE $c = \ cost[i][k]\ +\ cost[k][j]$
\STATE $cost[i][j] = cost[i][j]\ >\ c\ ?\ c\ :\ cost[i][j]$
\ENDFOR
\ENDFOR
\ENDFOR\\
\end{algorithmic}
\end{algorithm}

In Algorithm \ref{alg:ShortestPath}, line 17-25 is a typical Floyd algorithm for finding shortest paths between two vertexes, which can be leveraged directly to search for the best choice of fusion strategy for CNNs with no branch like VGG\cite{vgg} and YOLO\cite{yolov3}. To extend the algorithm to other computing graphs with branches, we set the operations that are dependent on more than one operation or by different operations to be barriers. We assume that fusion opportunities will never exist among operations via the barriers due to our fusion templates. Thus, we only need to determine the best choice of execution strategies among all pairs of barriers, as shown in Figure \ref{fig:processflow}.d. In Algorithm \ref{alg:ShortestPath}, $cost[i][j]$ denotes the execution time of operations from the $i^{th}$ node to the $j^{th}$ node.

Unfortunately, a problem may occur at the barriers while we directly leverage the Floyd algorithm in our study. For example, as shown in Figure \ref{fig:processflow}.c, if we fuse operations denoted by \textcircled{2} and \textcircled{5}, the edge $data1$ targeting $barrier1$, $data3$ targeting $barrier1$ in Figure \ref{fig:processflow}.d should be pruned as \textcircled{5} denotes an element-wise-add, which only needs to be implemented once. Additionally, horizontal fusion has a similar problem. In topologies such as Inception of GoogLeNet\cite{googlenet}, convolutions that share the same input feature maps can be fused to avoid reloading the input data. Once these convolutions are fused, other edges that proceed from the input feature map should be pruned. In our study, we enumerate these special cases at barriers (line 5-10). As shown in Figure \ref{fig:path-search}.c, baseline without operation fusion costs 2.137ms. By enumerating fusion opportunities at the bottom barrier in Figure \ref{fig:path-search}.d e, our heuristic algorithm finds the best choice of fusion strategy e for the subgraph with existing fusion templates.

\section{Evaluation}

This section proposes onboard evaluation to test the DNNVM and emphasize the performance improvement by operation fusion methods using devices for both embedded applications and data centres.

\subsection{Experiment Environment}

\begin{figure}[H]
\centering
\includegraphics[width=0.4\textwidth]{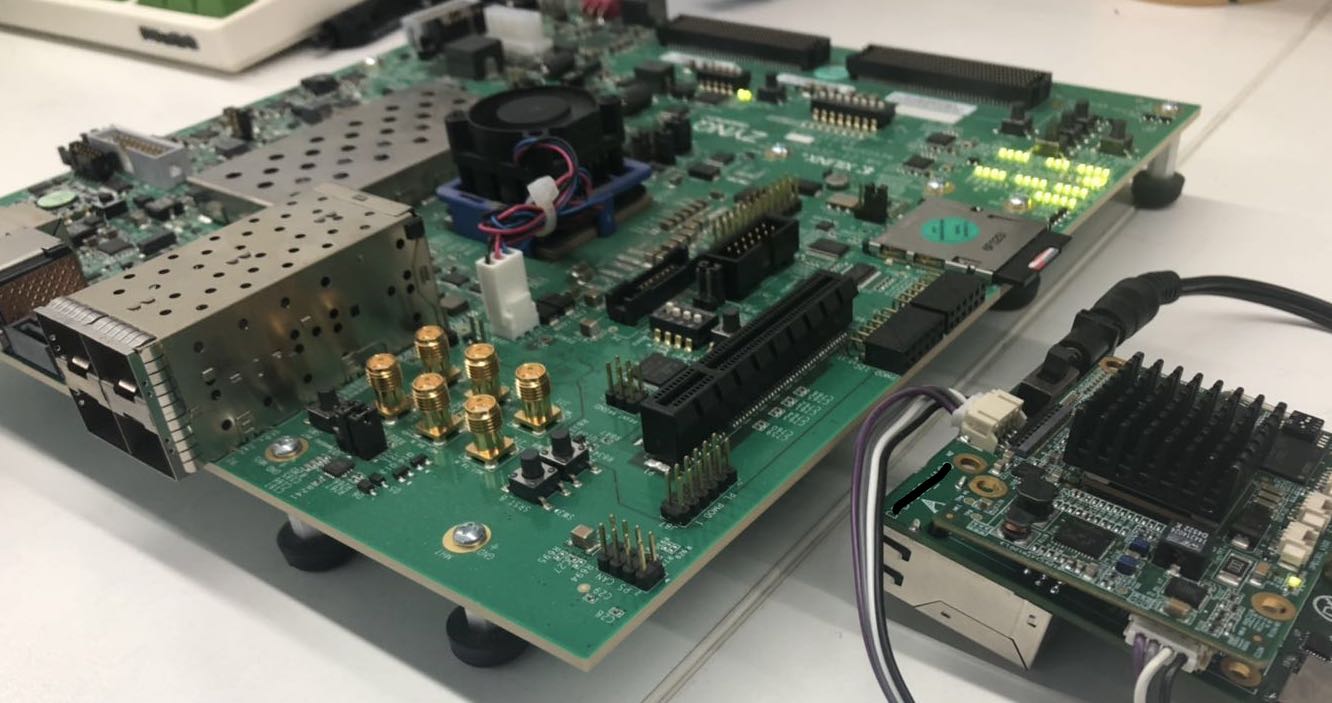}
\caption{ZU9(left) and ZU2(right) FPGA Devices}
\label{fig:platforms}
\end{figure}

As shown in Figure \ref{fig:platforms}, we employ Xilinx ZU2 (ic\_p = 24, oc\_p = 12, h\_p = 4) and ZU9 (ic\_p = 32, oc\_p = 16, h\_p = 8) FPGA devices to evaluate the DNNVM with operation fusion using our previously described designs. Our experiments are conducted on these devices. ZU2 is a low-cost FPGA chip that is commonly employed in embedded devices; it contains approximately 0.66 MB of on-chip storage. ZU9 has 4 MB of on-chip storage and targets data centre scenarios. Table \ref{tab:cost} shows our benchmark CNN models. Selected CNNs are extensively employed in multiple applications, including face recognition, object recognition, classification, and tracking. As a result, VGG \cite{vgg}, Resnet \cite{resnet}, and GoogLeNet \cite{googlenet} are selected for comparison with other designs. We train all neural network models with Caffe and perform adjustments to convert feature maps, weights and biases to 8-bit points from a 32-bit floating point. Our data quantization method is similar with Angel-eye\cite{angeleye}, the radix position of the fixed point data in each layer is chosen differently and we adopt the quantization method with the best accuracy by enumerating possible solutions. We map all operations, with the exception of fully connected layers, onto FPGA accelerators. We demonstrate that our design guarantees the practicality for both embedded platforms and data centres. We synthesize the hardware logic with Vivado 2017.1.

\subsection{Experimental Results}

The most critical characteristic of an FPGA-based accelerator is the achieved performance of the system. Peak performance reveals the optimized design of the hardware platform, but the comparison between peak performance of multiple designs is meaningless in some degree because the hardware never achieves the peak performance for specific CNN models in practice. So we implement typical CNN models and present achieved performance in practice. Firstly, as shown in Table \ref{tab:cost} and Table \ref{tab:comp}, our baseline design implementing instructions generated by DNNVM without pipeline optimizations on ZU2@330MHz. Baseline implementations have achieved appealing performance and the optimized implementations by leveraging operation fusion improves the throughput further.

\subsubsection{\textbf{Throughput Improvement by the Operation Fusion}}

\begin{figure*}
\centering
\includegraphics[width=0.9\textwidth]{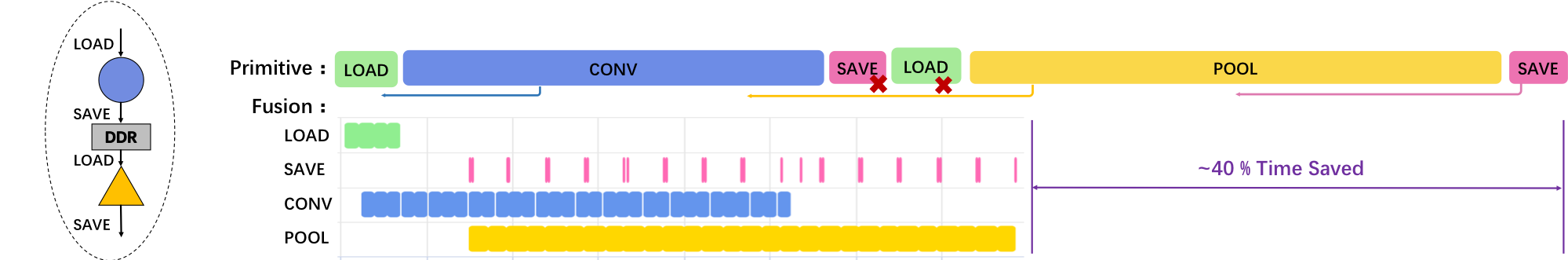}
\caption{Optimized pipeline achieved by fusing the adjacent convolution and pooling by DNNVM.}
\label{fig:profiling_cp}
\end{figure*}

\begin{figure*}
\centering
\includegraphics[width=0.9\textwidth]{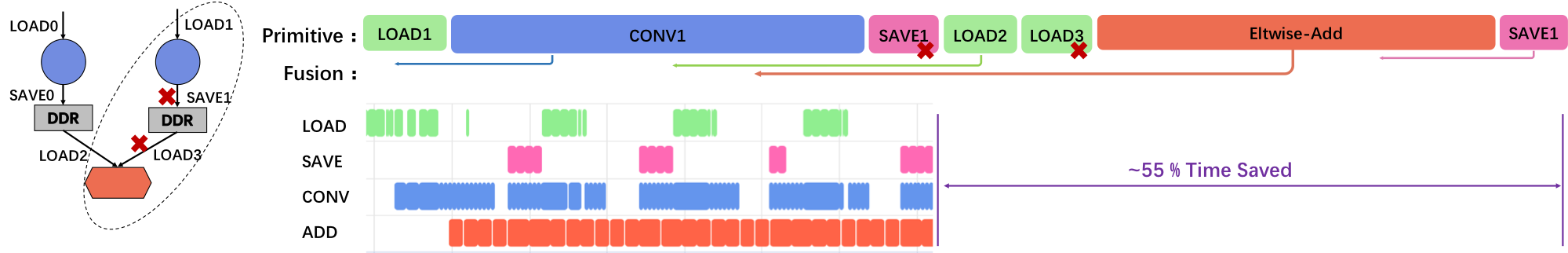}
\caption{Optimized pipeline achieved by fusing one of convolutions and the adjacent eltwise-add by DNNVM.}
\label{fig:profiling_resnet}
\end{figure*}

\begin{table*}[htbp]
    \centering
  \caption{Performance Improvement by Operation Fusion on ZU2@330MHz~(Peak Performance : 380GOPs/s)}
    \setlength{\tabcolsep}{1mm}{
    \centering
    \resizebox{\textwidth}{!}{
    \begin{tabular}{c|cccccc|cccc}
    \toprule
      \multirow{2}[0]{*}[-3pt]{\thead{Benchmark}} & \multicolumn{6}{c|}{\thead{Compilation Cost}} & \multicolumn{4}{c}{\thead{Results}}\\
       \cline{2-11}
     & \thead{Node\\Size} & \thead{Graph\\Generation(ms)} & \thead{Isomorphism\\Fusion(ms)} & \thead{Compilation\\Jobs(s)} & \thead{Evaluation\\(ms)} & \thead{Auto\\Tuning(ms)} & \thead{Baseline\\Perf(GOPs/s)} & \thead{Greedy~Fusion\\(GOPs/s)} & \thead{Optimized~Fusion\\(GOPs/s)} & \thead{Speedup}\\
        \hline
    \thead{VGG} & \thead{32}    & \thead{0.389} & \thead{9.28} &  \thead{28.266} & \thead{189.833} & \thead{0.647} & \thead{325.5} & \thead{331.5} & \thead{334} & \thead{1.03x}\\[-4pt]
    \thead{ResNet50} & \thead{120}   & \thead{4.112} & \thead{30.893} &  \thead{29.125} & \thead{55.097} & \thead{12.309} & \thead{195.4} & \thead{221.9} & \thead{228.7} & \thead{1.17x}\\[-4pt]
    \thead{ResNet152} & \thead{358}   & \thead{18.125} & \thead{145.023}  & \thead{72.151} & \thead{141.553} & \thead{70.53} & \thead{212.5} & \thead{233} & \thead{244.1} & \thead{1.15x} \\[-4pt]
    \thead{GoogLeNet} & \thead{137}   & \thead{3.424} & \thead{66.069}  & \thead{22.139} & \thead{28.076} & \thead{6.61} & \thead{183.1} & \thead{204.6} & \thead{231.5} & \thead{1.26x} \\[-2pt]
    \bottomrule
    \end{tabular}}
  \label{tab:cost}}
\end{table*}

To analyse the performance improvement by the operation fusion and pipeline optimizations, we test instructions generated by DNNVM on ZU2 @330 MHz. As shown in Figure \ref{fig:profiling_cp}, we fuse one adjacent convolution and pooling in GoogLeNet-v1\cite{googlenet}. The input feature maps are sliced to fit the capacity of on-chip memory. The intermediate data calculated by the convolution will not be stored to off-chip memory; thus, LOAD and SAVE between CONV and POOL can be skipped. As soon as an output feature map is achieved, SAVE is implemented which can be hidden in computations. In addition, LOAD, SAVE, CONV, POOL can be concurrently performed. We set the input feature map to be $28 \times 28 \times 32$, the convolution kernel size to be $5 \times 5$, the pooling kernel size to be $3 \times 3$, the stride of them to be 1 and the output feature map to be $28 \times 28 \times 256$. According to Equation \ref{eq:comp}, the total workloads are 0.32 GOPs, and according to Equation \ref{eq:ctcnof} \ref{eq:ctcf}, data transfer is 6.27 KB. On ZU2, 0.375 ms is needed for this convolution operation, and 0.242 ms is needed for the following pooling operation. The fusion methodology reduces the total data transfer by 64\% and achieves a 1.67x speedup. 

In Figure \ref{fig:profiling_resnet}, the element-wise-add operation needs to load the results from different convolution operations. The large amount of data exchange is one of the main types of overhead. As we fuse one adjacent convolution with the element-wise-add in Resnet50, SAVE 1 and LOAD 3 can be skipped. Tiling is used to fit the data into on-chip buffers. Communications and computations are implemented by different blocks concurrently. On ZU2, we need 0.467 ms and 0.463 ms for the convolutions, and an extra 0.833 ms is required for the element-wise-add. Fortunately, after fusing element-wise-add with one of the prior convolutions, the total execution time can be reduced to 1.039 ms. Fusion of convolution and element-wise-add achieves a 2.2x speedup and a 36.4\% reduction of data transfer compared with the primitive serial implementations.

Table \ref{tab:cost} shows the execution time of the compilation process and the on-board execution time of the generated instructions for entire CNN models. The results show that our operation fusion techniques outperform greedy algorithms and provide a speedup from 1.03$\times$ to 1.26$\times$ than na\"ive implementations generated by DNNVM. To further analyse how much the operation fusion can improve the resource utilization, we calculate the average performance normalized by the peak performance. In Table \ref{tab:cost}, the resource utilization rate (Achieved performance / Peak performance (380GOps/s)) increases by 10\% and 20\% on ResNet and GoogLeNet, respectively, but only approximately 2\% on VGG. Due to the variety of operations and complex topology in ResNet and GoogLeNet, operation fusion keeps the computation blocks busy and hides communication in computations, as shown in Figure \ref{fig:profiling_resnet}. However, VGG only has convolution and pooling and their original resource utilization is approximately 90\%; thus, the improvement is not impressive. Table \ref{tab:cost} indicates that compilation jobs and corresponding optimizations only cost dozens of seconds, which has a minimal impact on the entire deployment process. Our heuristic fusion algorithm can effectively improve the performance of complex neural networks with minimal compilation cost.

In our work, meaningful and fair comparisons require designs that leverage pipeline optimizations, especially operation fusion for the same CNN that targets the same FPGA device. It is hard to satisfied all requirements as mentioned, thus, we choose designs that leverage fusion technology on FPGA platforms as analogous as possible. We compare our design with them on either the hardware side or the compiler side. Additionally, the frequency of some hardware designs may be different and can not be improved due to the lack of hardware resources, so we compare our design with others at their highest frequency reported in the literature. 

\begin{table*}[htbp]
  \centering
  \caption{Overall Performance Comparisons to Other Designs Leveraging Operation Fusion}
  \label{tab:comp}
  \renewcommand{\multirowsetup}{\centering}
  \setlength{\tabcolsep}{0.7mm}{
  \resizebox{\textwidth}{!}{
    \begin{tabular}{c|ccccc|cccccc}
     \toprule \\[-13pt]
     \thead{} & \multicolumn{5}{c|}{\thead{Comparison to Fusion Designs}} & \multicolumn{6}{c}{\thead{Comparison to Compilers}} \\[-3pt]
             \hline \\[-11pt]
     \thead{Benchmark} & \thead{fpgaConvNet\cite{fpgaconvnet17}} & \thead{fuse1\cite{fuse1}} & \thead{fuse2\cite{fuse2}} & \thead{Ours} & \thead{Ours} & \thead{Snowflake\cite{snowflakecompiler}} & \thead{DnnWeaver\cite{dnnweaver}}& \thead{xfDNN\cite{xfdnn}} & \thead{DLA\cite{dla}} & \thead{Ours} & \thead{Ours} \\[-3pt]
             \hline \\[-11pt]
     \thead{Platform} & \thead{XC7Z045} & \thead{XC7Z045} & \thead{XC7Z045} & \thead{XC7Z020} & \thead{ZU2} & \thead{XC7Z045} & \thead{XC7Z020} & \thead{VU9P}  & \thead{Arria10 1150} & \thead{ZU9} & \thead{ZU9}\\[-3pt]
             \hline \\[-11pt]
     \thead{Batch} & \thead{1}& \thead{1}& \thead{1}& \thead{1}& \thead{1}& \thead{1} & \thead{1} & \thead{16}  & \thead{-} & \thead{1} & \thead{3}\\[-2pt]
             \hline \\[-11pt]
     \thead{Frequency(MHz)} & \thead{125} & \thead{100} & \thead{100} & \thead{200} & \thead{330} & \thead{250} & \thead{150} & \thead{500} & \thead{450} & \thead{500} & \thead{330}\\[-3pt]
             \hline \\[-11pt]
     \thead{On-chip memory(MB)} & \thead{2.4} & \thead{2.4} & \thead{2.4}& \thead{0.5}  & \thead{0.66} & \thead{2.4} & \thead{0.6} & \thead{8.9+35}  & \thead{-} & \thead{4} & \thead{4}\\[-3pt]
             \hline \\[-11pt]
     \thead{Arithmetic Precision} & \thead{16bit} & \thead{float32} & \thead{16bit} & \thead{8bit} & \thead{8bit} & \thead{16bit} & \thead{float32} & \thead{8bit} & \thead{minifloat} & \thead{8bit} & \thead{8bit}\\[-3pt]
             \hline \\[-11pt]
     \thead{VGG(OPs/s)} & \thead{155.8G} & \thead{162G} & \thead{230G} & \thead{204G} & \thead{334G} & \thead{-} & \thead{31.35G} & \thead{2.54T}  & \thead{-} & \thead{1.78T} & \thead{2.82T}\\[-3pt]
             \hline \\[-11pt]
     \thead{ResNet50(OPs/s)} & \thead{-}& \thead{-}& \thead{-}& \thead{130G}& \thead{228.7G}& \thead{35.2G} & \thead{-}  & \thead{1.29T} & \thead{-}& \thead{0.68T} & \thead{1.38T}\\[-3pt]
             \hline \\[-11pt]
      \thead{Googlenet(OPs/s)} & \thead{181G}& \thead{-}& \thead{-}& \thead{134G}& \thead{231.5G}& \thead{-} & \thead{-}  & \thead{2.07T} & \thead{2.8T} & \thead{0.7T} & \thead{1.41T}\\[-2pt]
    \bottomrule
    \end{tabular}}}
\end{table*}

\subsubsection{\textbf{Comparison to Fusion Designs}}

As shown in Table \ref{tab:comp}, fpgaConvNet \cite{fpgaconvnet17}, fuse1 \cite{fuse1} and fuse2 \cite{fuse2} optimize throughput on the hardware side, and fuse different operations into a single block. fpgaConvNet \cite{fpgaconvnet17} proposes the alternative exploitation of the capabilities of FPGAs and implements partitioning of a CNN along the depth into several subgraphs and then maps each subgraph into a different bitstream. Although reconfiguration overheads are added, fpgaConvNet achieves 48.53 GOPs/s for VGG on Zynq XC7Z020 and 155 GOPs/s on Zynq XC7Z045. Fuse1 \cite{fuse1} focuses on optimizing the external memory bandwidth utilization. This fused-layer accelerator reduces memory transfer from 77 MB to 3.6 MB for VGG. Unfortunately, this design requires 6.5$\%$ additional clock cycles with fusion. Fuse2 \cite{fuse2} explores algorithms to determine the fusion strategy for each layer; they explore fusion possibility by the branch and bound algorithm, depending on the hardware resources, bandwidth and workloads. Heterogeneous achieves 76.9 GOPs/s and 230 GOPs/s for Alexnet and VGG respectively. Table \ref{tab:util} provides the detailed utilization of hardware resources of our design on different platforms. We only use 25$\%$ BRAMs, 24$\%$ DSP, 33$\%$ FF and 14$\%$ LUT on ZU2 compared with the fusion design \cite{fuse2} on Zynq XC7Z045 but achieves 1.45x throughput and 1.8x energy efficiency even with the substantially smaller on-chip memory on ZU2. As leveraging fusion on the software side simplifies the hardware design, we can focus on optimizing frequency and resource utilization. In addition, we can more efficiently scale to deeper and more complicated CNNs by adjusting instructions with equivalent performance compared with these fusion designs.

\begin{table}[htbp]
\setlength{\abovecaptionskip}{0em}
  \tiny
  \centering
  \caption{Detailed Utilization of Hardware Resources}
  \setlength{\tabcolsep}{2mm}{
  \renewcommand{\arraystretch}{0.8}
  \begin{tabular}{c|cccc}
    \toprule \\[-8pt]
    & \thead{fuse1\cite{fuse1}} & \thead{fuse2\cite{fuse2}} & \thead{Ours} &  \thead{Ours}\\[-3pt]
    	\hline \\[-7pt]
    \thead{Platform}  & \thead{Zynq 7045} & \thead{Zynq 7045}  &  \thead{ZU2}   & \thead{ZU9}\\[-3pt]
        \hline \\[-7pt]
    \thead{BRAM(18k)}  & \thead{703}  & \thead{909}   & \thead{235}   & \thead{1494}\\[-3pt]
        \hline \\[-7pt]
    \thead{DSP}   & \thead{784}   & \thead{824}     & \thead{194}   & \thead{1542}\\[-3pt]
        \hline \\[-7pt]
    \thead{FF}    & \thead{90854} & \thead{120957}  & \thead{39597} & \thead{236970}\\[-3pt]
        \hline \\[-7pt]
    \thead{LUT}   & \thead{118400} & \thead{155886}  & \thead{21952} & \thead{117810}\\[-3pt]
        \hline \\[-7pt]
    \thead{Power(W)} & \thead{9.4} &  \thead{9.4}    &  \thead{7.5} & \thead{22.8}\\[-3pt]
        \hline \\[-6pt]
    \thead{Energy Efficiency\\(GOPs/s/W)} & \thead{17.25} & \thead{24.42} & \thead{44.5}   & \thead{123.7}\\[-2pt]
    \bottomrule
    \end{tabular}}
  \label{tab:util}%
\end{table}%

\subsubsection{\textbf{Comparison to Compilers}}

As shown in Table \ref{tab:comp}, Snowflake \cite{snowflakecompiler}, DnnWeaver \cite{dnnweaver}, VTA \cite{vta}(not in Table \ref{tab:comp}), xfDNN \cite{xfdnn}, and DLA \cite{dla} propose a FPGA-based design and optimize the implementations on the software side. Their compilers leverage fusion with a computing graph, loops, or co-optimizations. Snowflake and DnnWeaver provide a powerful compiler; however, their final performance cannot be guaranteed. VTA explores the simultaneous utilization of compute and memory resources by reducing the inference time of ResNet18 from more than 3 s to less than 0.5 s. Currently, xfDNN and DLA have shown a state-of-the-art performance. xfDNN adopts VU9P, which has 8.9 MB BRAMs and more than 35 MB UltraRAMs. As a result, xfDNN can pre-load a large number of parameters and feature maps onto on-chip memory to avoid the data exchange. Larger batch than 1 contributes to the performance improvement in xfDNN and more data re-usage of parameters. We outperform xfDNN on VGG and ResNet50. DLA is the most efficient acceleration on GoogLeNet. Unfortunately, we achieve a substantially lower performance than that of xfDNN and DLA on GoogLeNet. First, our frequency is limited to 330 MHz for batch 3 on ZU9 due to a lack of wiring resources. Second, GoogLeNet has many layers with small-scaled feature maps, frequent data exchange is caused by a substantially smaller on-chip memory than xfDNN and DLA, and cannot be hidden by fusion. Third, bandwidth saturation also causes a performance gap.

\section{Related Work}

Many domains in machine learning benefit from CNN algorithms. Due to the high computation complexity of a CNN, multiple compiler infrastructures occur, and various compiler technologies are employed to improve the throughput on GPPs or on specialized processors.

For GPP compilers, a side-effect free representation of operations, applicability and generality to different deep learning frameworks, and optimized scheduling are highlighted. Intel nGraph \cite{ngraph} and Google XLA \cite{xla} have the role of a bridge between deep learning frameworks and hardware back-ends. nGraph utilizes MKL-DNN to produce highly optimized implementations on CPUs and the PTX-emitting back-end of the LLVM to generate assembly code for GPUs. The XLA compiler acts as a back-end for TensorFlow. TVM \cite{tvm} proposes an ahead-of-time compiler that supports multiple front-ends and hardware platforms. These compilers adopt high-level computing graphs and leverage fusion across layers based on predetermined rules. Auto-scheduling algorithms have gradually attracted a substantial amount of attention and provide appealing productiveness. Tensor Comprehension \cite{tensorcomprehension} adopts polyhedral IRs TC and employs a genetic search of affine transformation options (e.g., tile sizes, loop fusion and scheduling strategies). PolyMage \cite{polymage} introduces fusion methods with loop nests and determines the rules of fusion and the range of tiling sizes to ensure a small auto-tuning space. Mullapudi et al. \cite{autoschedule} introduce an automatic fusion and tiling selecting method for Halide \cite{halide}. Jangda et al. \cite{tilefuse} develop a cost function for evaluating all valid fusion opportunities in only $O(n^{2})$ time instead of $O(2^{n})$. In this way, all potentially profitable fusion opportunities will considered compared with greedy algorithms \cite{polymage, autoschedule}.

Similarly, to accelerate a CNN on specialized accelerators, the expressive IRs and an optimized schedule to enhance data locality and parallelism are important. To leverage operation fusion and other pipeline strategies, several FPGA-based accelerators adopt a Streaming Architecture \cite{toolflow} and are optimized on the hardware side. Alwani et al. \cite{fuse1}, Xiao et al. \cite{fuse2} and fpgaConvnet \cite{fpgaconvnet17} instantiate a fusion design with full consideration of hardware resources. Although they achieve appealing performance for dedicated neural networks, the scaling of these designs to deeper networks is difficult on a resource-limited platform, and re-configuration overheads are introduced when switching to other models or applications.

Other accelerators leverage co-optimization upon hardware and software in a Single Computation Engine \cite{toolflow}. Compilers are designed to map NNs into instructions that are loaded on custom FPGA-based devices. These designs simplify the hardware design and usability of applications. Writing algorithms in compilers to maximize the hardware throughput can be more efficient. Additionally, instruction-based computation blocks can be reused for various operations. FP-DNN \cite{fpdnn} converts convolution into matrix multiplication and uses C++ for Host. However, communication strategies need to be carefully designed, a long compilation process is needed to realize these optimizations, even with high-level synthesis. VTA \cite{vta} together with TVM adopts fusion technology named Task-Level Pipeline Parallelism to generate optimized instruction chains. The main challenges of TVM/VTA lie in the combinatorial explosion of the fusion opportunities, and which fusion sequence is the optimal execution strategy has not been determined. Xilinx and Intel propose xfDNN \cite{xfdnn} and DLA \cite{dla}, respectively, as a tool chain to deploy CNN on custom accelerators and achieve the state-of-art throughput. They fuse many adjacent layers, and the weights and bias of these layers can be pre-loaded due to the large on-chip memory capacity. However, public evidence of their feasibility on platforms with less on-chip memory is not available.

\section{Conclusion}

In this study, we propose the end-to-end compiler infrastructure DNNVM. DNNVM is an integration of optimizers for framework-independent XGraph, loops and data layouts, an assembler, a runtime supporter and a validation environment. All modules in DMMVM are indispensable parts in a full-stack compiler design. Due to the high baseline throughput of hardware design, even one per cent performance improvement is hard to be achieved. We transform the optimization problems in compiler infrastructure, such as data layout and pipeline optimization, into several graph-level problems and leverage heterogeneous optimizations on the software side. Thus our flexible hardware structure with DNNVM takes advantage of both the software programmability and the efficiency of custom hardware. We achieve state-of-the-art performance over VGG and Resnet50 with much lower hardware resources consumption. On GoogLeNet with operation fusion, we achieve a maximum of 1.26$\times$ the speedup of na\"ive implementations. The hardware design principle and heterogeneous optimization algorithms (including all the fusion methods) can be extended to other FPGA-based CNN accelerators.

\renewcommand{\Large}{\fontsize{15pt}{\baselineskip}\selectfont}

\bibliographystyle{unsrt}
\bibliography{ref}

\end{document}